\documentclass[sigconf]{acmart}

\AtBeginDocument{%
  }

\usepackage{amsmath}
\usepackage{graphicx} 
\usepackage{microtype}
\usepackage{booktabs} 
\usepackage{multirow}
\usepackage{graphicx} 
\usepackage{float}
\usepackage{CJKutf8}
\usepackage{bookmark}
\usepackage{amsfonts}
\usepackage{amsmath}
\usepackage{adjustbox}
\usepackage[ruled,linesnumbered]{algorithm2e}
\usepackage{marvosym}
\copyrightyear{2023}
\acmYear{2023}
\setcopyright{rightsretained}
\acmConference[KDDCup '23]{To be presented at KDD Cup 2023 Workshop:
Multilingual Session Recommendation Challenge}{August 9, 2023}{Long Beach, CA, USA}
\acmBooktitle{KDD Cup 2023 Workshop: Multilingual Session Recommendation Challenge, August 9, 2023, Long Beach, CA, USA}
\acmDOI{}
\acmISBN{}

\begin{document}
\begin{CJK}{UTF8}{gbsn}

\title{Language-Enhanced Session-Based Recommendation with Decoupled Contrastive Learning}

\author{Zhipeng Zhang$^{*}$}
\email{zp.zhang@std.uestc.edu.cn}
\affiliation{%
  \institution{University of Electronic Science \\and Technology of China}
  \city{Chengdu}
  \country{China}
}

\author{Piao Tong$^{*}$}
\email{piaot@std.uestc.edu.cn}
\affiliation{%
  \institution{University of Electronic Science \\and Technology of China}
  \city{Chengdu}
  \country{China}
}

\author{Yingwei Ma$^{*}$}
\email{yingwei.ywma@gmail.com}
\affiliation{%
  \institution{National University of \\Defense Technology}
  \city{Changsha}
  \country{China}
}

\author{Qiao Liu\textsuperscript{\Letter}}
\email{qliu@uestc.edu.cn}
\affiliation{%
  \institution{University of Electronic Science \\and Technology of China}
  \city{Chengdu}
  \country{China}
}

\author{Xujiang Liu}
\email{575694112@qq.com}
\affiliation{%
  \institution{Pangang Group \\Research Institute}
  \city{Chengdu}
  \country{China}
}

\author{Xu Luo}
\email{39440938@qq.com}
\affiliation{%
  \institution{Pangang Group \\Research Institute}
  \city{Chengdu}
  \country{China}
}

\renewcommand{\authors}{Zhipeng Zhang, Piao Tong, Qiao Liu, Yingwei Ma∗, Xujiang Liu, Xu Luo}
\renewcommand{\shortauthors}{Zhipeng and Piao, et al.}




\begin{abstract}
Session-based recommendation techniques aim to capture dynamic user behavior by analyzing past interactions. However, existing methods heavily rely on historical item ID sequences to extract user preferences, leading to challenges such as popular bias and cold-start problems. In this paper, we propose a hybrid multimodal approach for session-based recommendation to address these challenges. Our approach combines different modalities, including textual content and item IDs, leveraging the complementary nature of these modalities using CatBoost. To learn universal item representations, we design a language representation-based item retrieval architecture that extracts features from the textual content utilizing pre-trained language models. Furthermore, we introduce a novel Decoupled Contrastive Learning method to enhance the effectiveness of the language representation. This technique decouples the sequence representation and item representation space, facilitating bidirectional alignment through dual-queue contrastive learning. Simultaneously, the momentum queue provides a large number of negative samples, effectively enhancing the effectiveness of contrastive learning. Our approach yielded competitive results, securing a 5th place ranking in KDD CUP 2023 Task 1. We have released the source code and pre-trained models associated with this work\footnote{\url{https://github.com/gaozhanfire//KDDCup2023}}.
\let\thefootnote\relax\footnotetext{*~~~The first four authors contributed equally to this work.}
\let\thefootnote\relax\footnotetext{\Letter~~Corresponding Author.}
\end{abstract}



\maketitle

\section{Introduction}

With the progress of data mining and machine learning, session-based recommendation, which employs customer session data to forecast their next click, is gaining momentum. Sequential recommender \cite{hidasi2015session, liu2018stamp} has garnered popularity owing to its capability of capturing users' transient and enduring preferences, rendering it widely applicable across diverse recommendation contexts.

Numerous methodologies have been proposed to model sequential recommendation scenarios, spanning from early matrix factorization approaches (e.g., FPMC \cite{rendle2010factorizing}) to more recent sequence neural networks (such as GRU4Rec \cite{hidasi2015session}, STAMP \cite{liu2018stamp}, and Transformer \cite{schifferer2022diverse}). While they may differ, their underlying principles are akin: capturing behavioral patterns that mirror user preferences based on the sequential access structure and establishing a mapping between user preferences and specific items grounded in their interaction relationships. However, the dependence on item interaction relationships curtails their efficacy when faced with the cold-start predicament of insufficient user interactions with certain items.

Recent research has explored the use of natural language text, such as item titles and descriptions, to gain domain knowledge and address the challenges mentioned earlier \cite{hou2022towards, li2023text}. This approach involves utilizing pre-trained language models like BERT to obtain text representations and learning how to transform these representations into item representations. While previous attempts have shown promise, there are still unresolved challenges that need to be addressed: \begin{figure}[htbp]
  \centering 
  \includegraphics[width=0.47\textwidth]{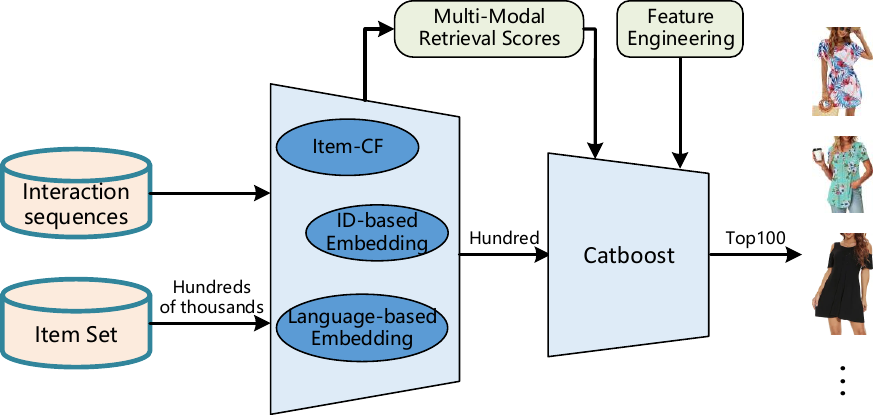}
  \caption{The Holistic Framework of the Proposed Adaptive Hybrid Multimodal Approach.}
  \label{fig:figure_1}\vspace{-2em}
\end{figure}(1)\textit{Insufficient integration of multi-modal information}: Existing methods \cite{hou2022towards} combine the text representation of an item with its corresponding item representation, but this approach can introduce interference between modalities. For example, if an item has dense interaction data and the item embedding's representation capability surpasses that of the text representation, incorporating the text representation may negatively impact the expressive power of the item representation. (2)\textit{Partial Semantic Synchronization}: Current approaches often treat candidate items as targets, resulting in parameter updates that only align the sequential representation with item representations that match sequence preferences. As a result, the alignment between item representations and sequential representations is considered, while the alignment in the reverse direction is overlooked.

In order to address the aforementioned challenges, we propose an adaptive hybrid multimodal framework for session-based recommendation. 
Figure \ref{fig:figure_1} provides an illustrative overview of the framework. This innovative framework employs multiple distinct models to encode autonomous representations of item IDs and their corresponding text sequences derived from a user's historical items. These representations are then leveraged to generate separate predictions for the subsequent click. To achieve dynamic fusion of the predictions, we employ the CatBoost algorithm, leveraging its capability to compute the conditional probability distribution of a class based on specific feature conditions. Additionally, we introduce Decoupled Contrastive Learning to enhance the representation capability of item text. This method separates the negative sample space into two independent subspaces: sequential representation and item representation. By decoupling these two spaces and utilizing a dual-momentum queue \cite{chen2020improved}, a substantial number of negative samples are updated and stored. Two types of contrastive learning losses are then computed. This process ultimately enables bidirectional deep alignment between sequential representations and item representations. Furthermore, our team actively participated in the KDD CUP 2023, where we conducted extensive evaluations of our proposed solution using the Amazon-M2 dataset \cite{jin2023amazon}. Through our rigorous experiments, we achieved competitive results, securing a  5th place ranking in task 1.
\section{METHODOLOGY}
\subsection{Multimodal Approach to Item Retrieval}
\subsubsection{ItemCF}
Despite the remarkable success of deep-learning-based approaches in recommender systems, traditional methods like collaborative filtering (CF) continue to find extensive applications on online platforms, owing to their irreplaceable strengths. These strengths include interpretability, minimal reliance on hardware, and high efficacy in both training and deployment.

In this competition, we incorporated several crucial factors into the construction of the item similarity matrix, including position distance information, directional cues in interactions, and normalizations based on item popularity. Formulas 1-4 encapsulate the fundamental concepts we employed to achieve these outcomes.

\begin{equation}
\operatorname{sim}(x, y)=\sum_{x, y \in Session} \frac{W_{\text {dist }} * W_{\text {dire }}  * W_{\text {pos}} }{|x|^{0.8} * |y|^{0.15}}
\end{equation}

\begin{equation}
W_{dist}(x, y)=\frac{1}{\left(\mid \operatorname{Pos}_x-\text { Pos }_y \mid+1\right) ^2}
\end{equation}

\begin{equation}
W_{dire}(x, y)= \begin{cases}1 & \text { if } {Pos}_x < {Pos}_y \\ 1/3 & \text { if } {Pos}_x > {Pos}_y\end{cases}
\end{equation}

Our research presents a unique approach by incorporating positional weights. It stems from the observation that the last item clicked by users, after multiple interactions, often reflects their meticulously chosen ultimate preference.
\begin{equation}
W_{pos}(x, y)= \begin{cases}1.8 & \text { if } {Pos}_y = |Session| - 1 \\ 1 & \text { if } {Pos}_y < |Session| - 1\end{cases}
\end{equation}

\subsubsection{ID-based Embedding : GRU}
\begin{figure}[htbp]
  \centering 
  \includegraphics[width=0.47\textwidth]{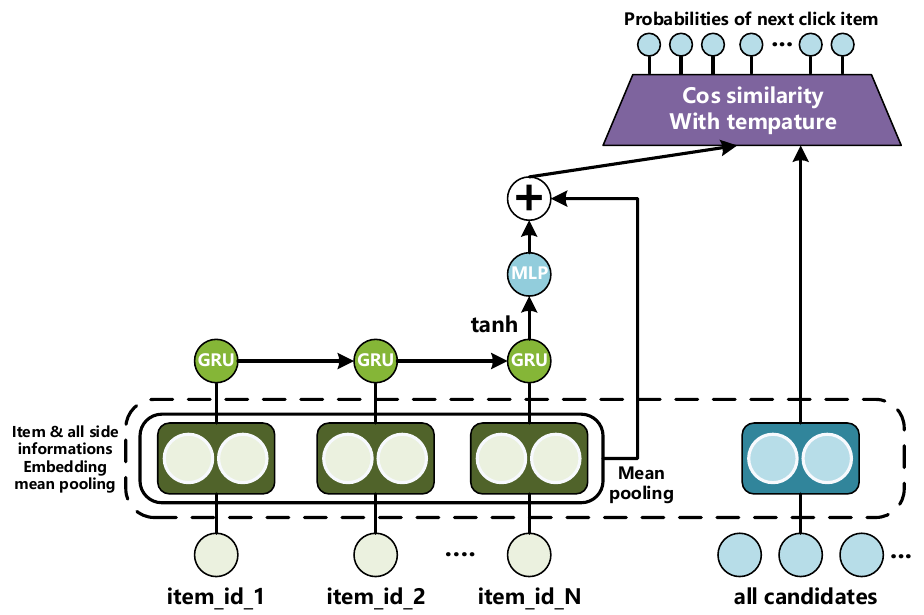}
  \caption{The customized GRU architecture.}
  \label{fig:figure_2}
\end{figure}

Recurrent neural networks are the most popular choice for modeling sequential data. We designed a customized RNN which achieves a 0.3776 public leaderboard score. The model architecture is shown in Figure \ref{fig:figure_2}. 

During the phase of shared embedding lookup on the user's last N clicked items, we utilize mean pooling on the embeddings of the items as well as their associated side information (such as price and brand) to derive fused embeddings. Subsequently, these embeddings are fed into a GRU layer for the extraction of sequential features. To address the challenge of vanishing gradients, we introduce a sequence-level residual connection block. This block conducts mean pooling on the initial fused embeddings prior to the GRU layer. It is then added to the output of the GRU layer to generate the final sequential representation. Differing from the approach adopted in \begin{figure*}[htbp]
  \centering 
  \includegraphics[width=0.7\textwidth]{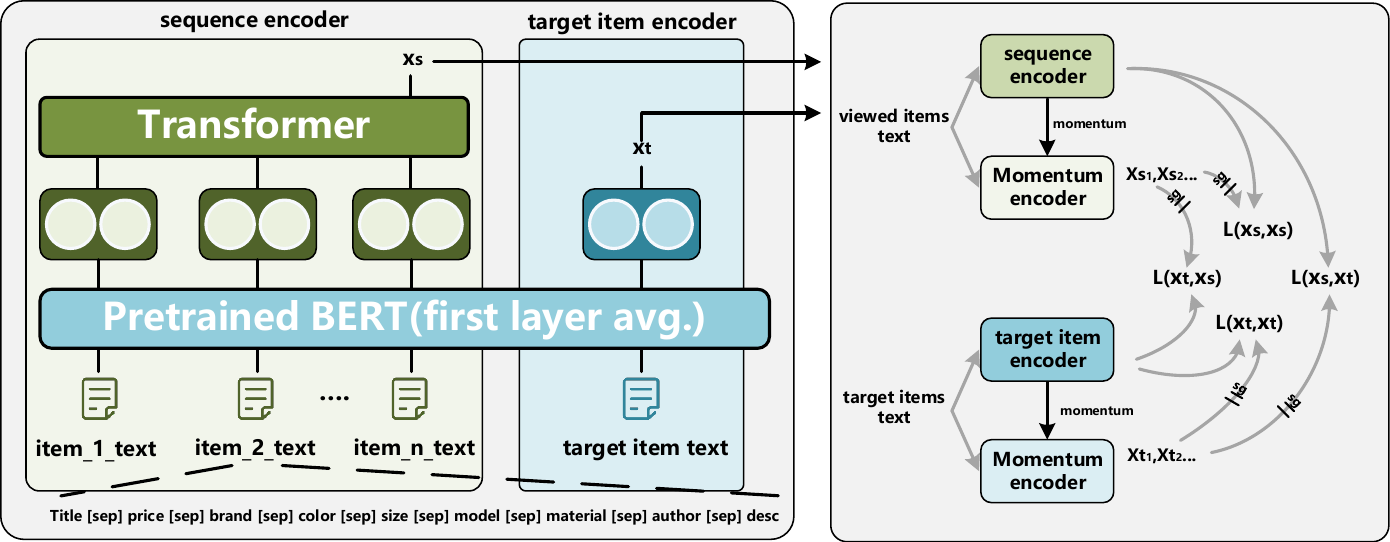}
  \caption{Visualization of Language Embedding Generation and Training Flow of Decoupled Contrastive Learning.}
  \label{fig:figure_3}\vspace{-1em}
\end{figure*}
previous studies \cite{schifferer2022diverse}, which involves mean-pooling the GRU layer outputs at each time step, our method preserves the
original fused embeddings, ensuring facilitating smoother gradient propagation.
\subsubsection{Language Embedding}
Contrastive learning has found wide applications in information retrieval problems \cite{gao2021simcse, ma2023mulcs}. Recently, its potential has also been explored in the field of session-based recommendation \cite{li2023text}. Motivated by these studies, this paper proposes a vector-based approach for item representation. The model architecture is shown in Figure \ref{fig:figure_3}.We segment the attribute texts of items using [SEP] and encode them using a pre-trained BERT model. This encoding process includes both previously browsed items and the target item. Finally, we perform average pooling on the outputs of the first layer and apply L2 normalization to obtain the representation vector for each item.

Each BERT encoder has a momentum encoder, whose parameters $\theta$ are updated by exponential moving average of the base encoder as follows:
$\boldsymbol{\Theta }_t \leftarrow \beta \boldsymbol{\Theta}_{t-1}+(1-\beta) \boldsymbol{\Theta}_{\text {base }}$. In each iteration step t, we maintain two memory queues: one for sequence representations and the other for target item representations. These queues store $K$ vectors, which are encoded by the momentum encoder from the most recent batches. With each optimization step, the oldest vectors in the memory queues are replaced by the vectors from the current batch. The momentum coefficient $\beta \in [0,1]$, usually set close to 1 (e.g., 0.999), ensures consistency across batches. Additionally, setting $K$ to a large value (>10000) ensures an abundant supply of negative samples, thereby promoting the learning of robust representations.

To train the encoder, we employ the ArcCon Loss \cite{zhang2022contrastive}, a loss function that introduces a hyperparameter $m$ to add a decision boundary, promoting compactness among sentence representations with the same semantics and increasing the dissimilarity between representations of sentences with different semantics:
\begin{equation}
\mathcal{L}(x,y)=-\log \frac{e^{s\cos \left(\theta_{x, y_{i}}+m\right)}}{e^{s\cos \left(\theta_{x, y_{i}}+m\right)}+\sum_{j \neq i} e^{s\cos \left(\theta_{x, y_{j}}\right) }}
\end{equation}
In Equation 5, the variable $s$ denotes the temperature hyperparameter. In essence, Equation 5 can be perceived as a (K+1)-way softmax classification, where $y = y_i$ represents the positive sample, and the items stored in the queue $\{y_i\}_{i\ne j}^{K}$ represent the negative samples. It is important to note that gradients do not propagate through the momentum encoder or the storage queue during backpropagation.

An important contribution in our study is the introduction of the training method called Decoupled Contrastive Learning. We divide the set of negative samples into two separate subsets, each containing all sequential representations and target item representations, respectively. By doing so, we can construct a symmetrical contrastive learning loss with the following form:
\begin{equation}
min\ \lambda_1\mathcal{L}(x_s,x_t) + \lambda_2\mathcal{L}(x_t,x_s)+\lambda_3\mathcal{L}(x_s,x_s)+\lambda_4\mathcal{L}(x_t,x_t)
\end{equation}

The hyperparameter $\lambda$ serves to harmonize the losses of each task. Upon completion of the training process, we discard the momentum encoders and memory queues, retaining solely the base encoders for the computation of sentence representations.
\subsection{Multimodal Fusion Rerank}

In this section, we retrieve items for each session using three retrieval methods, resulting in three sets of retrieved items. To combine the scores from the retrieval methods, we multiply them together, represented as $y_i = \prod_{t=1}^{3}y_t$, where $y_t$ is the score for each retrieval method. From the fused scores, we select the top 120 items as the preliminary retrieval results. Next, we integrate the scoring of retrieval results from different modalities and perform fine-grained feature engineering. We employ the CatBoost algorithm, which is based on decision trees and can effectively merge the retrieval results from different modalities by adaptively combining features.
\subsubsection{Item Hot Features}
The fundamental characteristics of item view frequencies in the dataset are considered. We have also incorporated the sort orders feature to enhance the model's ability to learn the ranking task with greater precision.
\subsubsection{Session Features}
In harmony with item features, we conduct composite statistics based on the item characteristics that users engage with within each session. This includes computing aggregate metrics such as the mean popularity and average price of all items within a session.
\subsubsection{Graph Features}
This study constructs a co-occurrence graph based on itemcf's co-occurrence relationship matrix. Each item is a node in the graph, and the weights between items serve as edge weights. By examining the degree relationships between items and their neighboring nodes, these features describe the importance and connectivity of candidate items within the co-occurrence graph. Our analysis includes node importance measures like PageRank, centrality features including betweenness centrality, Katz centrality, and degree centrality, as well as neighbor node relationship characteristics such as mean, count, maximum, and standard deviation of edge weights. Integrating these features enhances the model's understanding of candidate item characteristics.

\section{Experiments}
\subsection{Datasets}
We applied data augmentation techniques referenced in the literature [3,8] to enhance the Amazon-M2 dataset across the UK, DE, and JP regions. Through a five-fold cross-validation method, the data was partitioned into training and validation sets. Notably, the validation set exclusively comprised sessions without data augmentation, with the final item serving as the label. The validation set not only served for evaluating the model's performance but also facilitated training set generation during the reranking stage. To ensure efficiency, the validation was conducted exclusively on the UK region's data.

\subsection{Overall Performance}
Our approach achieved considerably performance gain over the baseline solution, and ranked top-5 in task-1.
The main results are shown in Table~\ref{tab:table_1}. 

\begin{table}[h]

  \centering \begin{tabular}{lcc}
\specialrule{0.7pt}{0pt}{2pt}
\textbf{Dataset} & \multicolumn{1}{l}{\textbf{Metric}} & \multicolumn{1}{l}{\textbf{Ranking}} \\ \specialrule{0.4pt}{0pt}{2pt}
leaderboard        & mrr@100=0.4033                      & 5th                                  \\
validation       & mrr@100=0.3394                      & -                     \\ \specialrule{0.7pt}{1pt}{2pt}
\end{tabular}
  \caption{Performance of our approach.}
  \label{tab:table_1} 
  \vspace{-3.0em}
\end{table}

\subsection{Ablation Studies}
We have delved into exploring the impact of applying Decoupled Contrastive Learning on model performance. Table~\ref{tab:table_2} presents the performance of the model in terms of the MRR@100 metric after three epochs of training.It is important to note that the default hyperparameters utilized in this study were set as 
$\lambda_1$:$\lambda_2$:$\lambda_3$:$\lambda_4$=0.35:0.35:
0.15:0.15.
\begin{table}[h]
\centering
\begin{tabular}{lcc}
\specialrule{0.7pt}{1pt}{2pt} 
\multicolumn{1}{c}{\textbf{Variants}}                                & \textbf{CV-MRR@100} \\
\specialrule{0.4pt}{0pt}{2pt} 
w/o Decoupled Contrastive Learning                                   & 0.2610              \\
$\lambda_1$:$\lambda_2$:$\lambda_3$:$\lambda_4$=0.5:0.5:0:0          & 0.2687              \\
$\lambda_1$:$\lambda_2$:$\lambda_3$:$\lambda_4$=0.25:0.25:0.25:0.25  & 0.2676              \\
$\lambda_1$:$\lambda_2$:$\lambda_3$:$\lambda_4$=0.45:0.3:0.125:0.125 & 0.2718              \\
\specialrule{0.3pt}{0pt}{2pt}
Language Emb                                                         & \textbf{0.2725}              \\
\specialrule{0.7pt}{1pt}{2pt} 
\end{tabular}
\caption{The influence of Decoupled Contrastive Learning.}
\label{tab:table_2}
\vspace{-2.0em}
\end{table}

We observed that by decoupling within the negative sample space and employing symmetric contrastive loss, the model's performance can be enhanced by 0.011. Additionally, adjusting the weight ratios of each loss function also influences the model's effectiveness.

As illustrated in Table~\ref{tab:table_3}, we evaluated the performance of the Multimodal Fusion Rerank methodology. Initially, we presented the performance metrics of various standalone models for comparative purposes against Multimodal Fusion. The experimental findings demonstrate that incorporating the catboost algorithm and conducting meticulous feature engineering both contribute significantly to elevating the performance of Multimodal Fusion.
\begin{table}[h]
\centering
\begin{tabular}{lcc}
\specialrule{0.7pt}{0pt}{2pt}
\textbf{Variants} & \textbf{CV-MRR@100} & \textbf{Leaderboard} \\
\specialrule{0.4pt}{0pt}{2pt}
GRU               & 0.3073              & 0.3776               \\
ItemCF            & 0.2941              & -                    \\
Language Emb      & 0.2725              & -                    \\
\specialrule{0.1pt}{0pt}{2pt} \specialrule{0.3pt}{0pt}{2pt}
Catboost Fusion   & 0.3175              & -                    \\
++Feature Engineering & \textbf{0.3394}          & \textbf{0.4034}               \\
\specialrule{0.7pt}{0pt}{2pt}
\end{tabular}
\caption{Performance of the Multimodal Fusion Rerank.}
\label{tab:table_3}
\vspace{-2.6em}
\end{table}

\section{Conclusion and Future Work}
This paper introduces a hybrid multimodal recommendation system that combines itemID and language modalities for item retrieval. We incorporate the catboost algorithm and employ fine-grained feature engineering to merge the retrieval results from both modalities. Our future work includes investigating the applicability of our algorithm in real-world e-commerce scenarios at a large scale.
\vspace{-0.3em}
\section*{ACKNOWLEDGMENTS}
This work is jointly supported by the National Natural Science Foundation of China (U19B2028), the National Science and Technology Major Project of the Ministry of Science and Technology of China (2022YFB4300603), the Sichuan Science and Technology Program (2023YFG0151) and the Development of a Big Data-based Platform for Analyzing the Coupling Relationship of Strip Production Processes Project.
\bibliographystyle{ACM-Reference-Format}
\bibliography{custom-base-leaderboardcar}

\end{CJK}
\end{document}